\documentclass[aps,prb,twocolumn,superscriptaddress,showpacs,superbib,floats]{revtex4}
\usepackage{graphicx}
\usepackage{amssymb}
\usepackage{amsmath}
\usepackage{dcolumn}
\usepackage{color}
\usepackage{multirow}
\begin{document}

\title{Quasi-one-dimensional antiferromagnetism and multiferroicity in CuCrO$_4$}

\author{J. M. Law}
\email{j.law@fkf.mpg.de}
\affiliation{Max-Planck-Institut f\"ur Festk\"orperforschung,
Heisenbergstr. 1, D-70569 Stuttgart, Germany}
\affiliation{Univ. Loughborough, Dept. Phys., Loughborough LE11 3TU, Leics. England }

\author{P. Reuvekamp}
\affiliation{Max-Planck-Institut f\"ur Festk\"orperforschung,
Heisenbergstr. 1, D-70569 Stuttgart, Germany}

\author{R. Glaum}
\affiliation{Institut f$\ddot{u}$r Anorganische Chemie, Universit$\ddot{a}$t Bonn, Gerhard-Domagk-Strasse 1, 53121 Bonn, Germany}

\author{C. Lee}
\affiliation{Department of Chemistry, North Carolina State
University, Raleigh, North Carolina 27695-8204, U.S.A.}

\author{J. Kang}
\affiliation{Department of Chemistry, North Carolina State
University, Raleigh, North Carolina 27695-8204, U.S.A.}

\author{M.-H. Whangbo}
\affiliation{Department of Chemistry, North Carolina State
University, Raleigh, North Carolina 27695-8204, U.S.A.}

\author{R. K. Kremer}
\affiliation{Max-Planck-Institut f\"ur Festk\"orperforschung,
Heisenbergstr. 1, D-70569 Stuttgart, Germany}

\date{\today}


\begin{abstract}

The bulk magnetic properties of the new quasi-one-dimensional Heisenberg antiferromagnet, CuCrO$_4$, were characterized by magnetic susceptibility, heat capacity, optical spectroscopy, EPR and dielectric capacitance measurements and density functional evaluations of the
intra- and interchain spin exchange interactions.
We found type-II multiferroicity below the N\'{e}el temperature of 8.2(5)~K, arising from competing antiferromagnetic nearest-neighbor ($J_{\rm nn}$) and next-nearest-neighbor ($J_{\rm nnn}$) intra-chain spin exchange interactions. Experimental and theoretical results indicate that the ratio
$J_{\rm nn}$/$J_{\rm nnn}$ is close to 2, putting CuCrO$_4$  in the vicinity of the
Majumdar-Ghosh point.

\end{abstract}

\maketitle

\subsection{Introduction}

Ferroelectricity driven by magnetic ordering in so-called type-II multiferroics
exhibits a high potential for technological applications. Switching  ferroelectric polarization by a magnetic field or magnetization by an electric field
offers  unprecedented applications in modern energy-effective electronic data storage technology.\cite{Auciello1998,Spaldin2010}
However, the link of magnetic order and ferroelectricity in type-II multiferroics still remains an intriguing question.\cite{Mostovoy2006,Cheong2007,Khomskii2009,Tokura2010}
To elucidate this issue, lately much attention  has been focused on the magnetic and magnetoelectric (ME) properties of quasi-one-dimensional (1D) antiferromagnetic (afm) quantum chain systems, which exhibit incommensurate cycloidal magnetic ordering.\cite{Brink2008} Such systems lose inversion symmetry and appear to be suitable candidates for multiferroicity.
Incommensurate spin-spiral magnetic ordering occurs in magnetic systems consisting of 1D chains when
the intrachain  nearest-neighbor (nn) and next-nearest-neighbor (nnn)  spin exchange interactions ($J_{\rm {nn}}$ and $J_{\rm {nnn}}$, respectively) are spin-frustrated, as found for compounds with CuX$_2$ ribbon chains made up of CuX$_4$ plaquettes, where X is a suitable anion, e.g. oxygen or a halide. Current examples include LiCuVO$_4$, NaCu$_2$O$_2$, CuCl$_2$.\cite{Gibson2004, Enderle2005, Capogna2005, Kimura2008, Banks2009, Capogna2010, Mourigal2011} It is typical that the Cu-X-Cu superexchange $J_{\rm {nn}}$ is ferromagnetic (fm), the Cu-X$\ldots$X-Cu super-superexchange $J_{\rm {nnn}}$ is afm and larger in magnitude.\cite{Banks2009,Koo2011}
A cycloidal spin-spiral along a 1D chain induces a macroscopic electric polarization,
$\vec{P}$ $\propto$ $\vec{e}_{ij} \times (\vec{S}_i\times \vec{S}_j)$,
where $e_{ij}$ is the vector linking the moments residing on adjacent spins $\vec{S_{i}}$ and $\vec{S_{j}}$.\cite{Katsura2005, Sergienko2006, Xiang2007}

In an ongoing effort to identify new quantum spin chain systems which potentially exhibit spiral magnetic order and ferroelectric polarization, we recently focused our attention on compounds crystallizing with ribbon chains, mainly those belonging to the CrVO$_4$ structure-type. The aforementioned structure-type features MO$_2$ ribbon chains where M is a magnetic 3$d$ transition metal. Such compounds were recently shown to exhibit exotic magnetic ground-states.\cite{Attfield1985,Glaum1996,Law2010,Law2011} Here, we report on the magnetic and ME properties of another member of this structure-type, CuCrO$_4$.
Our density functional calculations indicate $J_{\rm {nn}}$ to be about twice as strong as $J_{\rm {nnn}}$ putting CuCrO$_4$ in the vicinity of the Majumdar-Ghosh point for which the ground state can by exactly solved.\cite{Majumdar1969} This feature makes CuCrO$_4$ uniquely exceptional since all of the CuX$_2$ ribbon chain systems investigated so far exhibit fm $J_{\rm {nn}}$ and afm $J_{\rm {nnn}}$ spin exchange, where $J_{\rm {nnn}}$ is considerably larger in magnitude than $J_{\rm {nn}}$.\cite{Enderle2005, Enderle2010, Banks2009, Koo2011} We demonstrate that CuCrO$_4$ exhibits long-range afm ordering below  $\sim$~8.2~K, which is accompanied by a ME anomaly due to possible spin-spiral ordering in the CuO$_2$ ribbon chains.

\section{Crystal Structure}\label{Structure}

\begin{figure}
  \includegraphics[width=7.5cm]{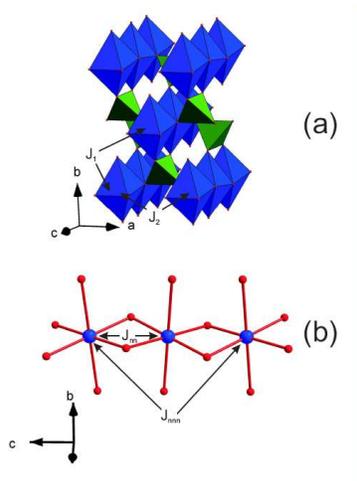}\\
  \caption{(Color online)(a): The crystal structure of CuCrO$_4$. The (blue) octahedra are the CuO$_6$ units while the (green) tetrahedra are the CrO$_4$ units. The interchain spin exchange pathways $J_{1}$ and $J_{2}$ are also indicated. (b): A section of the CuO$_2$ ribbon chain highlighting the edge sharing CuO$_4$ plaquettes, with the nn $J_{\rm {nn}}$ and nnn $J_{\rm {nnn}}$ spin exchange pathways labeled.}\label{Fig1}
\end{figure}

CuCrO$_4$ crystallizes in the CrVO$_4$ structure-type\cite{Brandt1943,Seferiadis1986} (SG: \textit{Cmcm}, No. 63) with Cu$^{2+}$ (\emph{d}$^9$, $S$~=1/2) and Cr$^{6+}$ (\emph{d}$^0$) ions. In the crystal structure of CuCrO$_4$, the axially-elongated CuO$_6$ octahedra share edges to form chains running along the $c$-axis (Fig. \ref{Fig1}(a)). These chains are interconnected by CrO$_4$ tetrahedra such that each CrO$_4$ tetrahedron is linked to three CuO$_4$ chains by corner-sharing (Fig. \ref{Fig1}(b)). The x$^2$-y$^2$ magnetic orbital of each CuO$_6$ octahedron is contained in the CuO$_4$ plaquette with four short Cu-O bonds.\cite{Whangbo2003} Thus, as far as the magnetic properties are concerned, CuCrO$_4$ consists of corrugated CuO$_2$ ribbon chains running along the $c$-axis (Fig. \ref{Fig1}(a)). At room temperature the Cu$\ldots$Cu distance is 2.945(2)~\AA\ and  the Cu-O-Cu $\angle$ is 98.1(1)$^{\rm o}$.

\section{Spin Exchange Interactions}\label{SecTheory}

\begin{figure}[htp]
\includegraphics[width=8cm ]{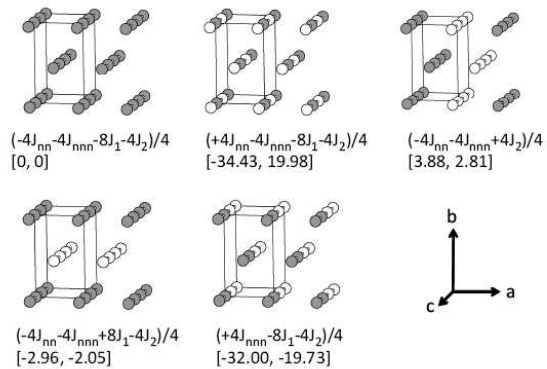}
\caption{Five ordered spin states used to extract the values of $J_{\rm {nn}}$, $J_{\rm {nnn}}$, $J_1$, and $J_2$, where the Cu$^{2+}$ sites with different spins are denoted by filled and empty circles. For each ordered spin state, the expression for the total spin exchange energy per 4 FUs is given, and  the two numbers in square bracket (from left to right) are the relative energies, in meV per 4 FUs, obtained from the GGA+$U$ calculations with $U_{\rm eff}$ = 4 and 6 eV, respectively.}
\label{Fig2}
\end{figure}

To examine the magnetic properties of CuCrO$_4$, we consider the four spin exchange paths defined in Fig. \ref{Fig1}; the two intra-chain exchanges $J_{\rm {nn}}$ and $J_{\rm {nnn}}$ as well as the inter-chain exchanges $J_{1}$ and $J_{2}$. To determine the values of $J_{\rm {nn}}$, $J_{\rm {nnn}}$, $J_{1}$ and $J_{2}$, we examine the relative energies of the five ordered spin states depicted in Fig. \ref{Fig2} in terms of the Heisenberg spin Hamiltonian,

\begin{equation}\label{eq1}
H = -\sum J_{ij} \vec{{S_i}}\vec{{S_j}},
\end{equation}

where $J_{\rm ij}$ is the exchange parameter (i.e., $J_{\rm {nn}}$, $J_{\rm {nnn}}$, $J_{1}$ and $J_{2}$) for the interaction between the spin sites $i$ and $j$. Then, by applying the energy expressions obtained for spin dimers with \textit{N} unpaired spins per spin site (in the present case, \textit{N} = 1),\cite{Dai} the total spin exchange energies of the five ordered spin states, per four formula units (FUs), are given as summarized in Fig. \ref{Fig2}. We determine the relative energies of the five ordered spin states of CuCrO$_4$ on the basis of density functional calculations with the Vienna \textit{ab initio} simulation package, employing the projected augmented-wave method, \cite{Kresse1993,Kresse1996a,Kresse1996b} the generalized gradient approximation (GGA) for the exchange and correlation functional,\cite{Perdew1996} with the plane-wave cut-off energy set to 400 eV, and a set of 64 \textbf{k}-points for the irreducible Brillouin zone. To account for the strong ele
 ctron correlation associated with the Cu 3$d$ state, we performed GGA plus on-site repulsion (GGA+$U$) calculations with $U_{\rm eff}$ = 4 and 6 eV for Cu.\cite{Dudarev1998} The relative energies of the five ordered spin states obtained from our GGA+$U$ calculations are summarized in Fig. \ref{Fig2}. Then, by mapping these relative energies onto the corresponding relative energies from the total spin exchange energies,\cite{Whangbo2003,Koo2008a,Koo2008b,Kang2009,Koo2010} we obtain the values of the spin exchange parameters, $J_{\rm {nn}}$, $J_{\rm {nnn}}$, $J_1$, and $J_2$ as summarized in Table \ref{Table1}.

\begin{table}[tbh]
\begin{ruledtabular}
\begin{tabular}{cccccc}
  $J_i$ & \multicolumn{2}{c}{$U_{\rm eff}$ = 4 eV} &  \multicolumn{2}{c}{$U_{\rm eff}$ = 6 eV} & experiment\\
\hline
$J_{\rm {nn}}$ & -199.7(1.0)& -63.8 & -115.9(1.0)& -55.4 & -54 \\ \\
$J_{\rm {nnn}}$ & -85.8(0.43)& -27 & -56.5(0.49) &-27 &-27  \\ \\
$J_1$ & -8.6(0.04)& -2.7 & -6.00(0.05) & -2.9 &-  \\ \\
$J_2$ & +31.1(0.16)& +9.8 & +22.3(0.19)& +10.7& +12 \\
\\
$\theta _{\rm CW}$& & -43.2&&-38.8&-56/-60\\
\end{tabular}
\end{ruledtabular}
\caption[]{Spin exchange parameters $J_{\rm {nn}}$, $J_{\rm {nnn}}$ $J_1$ and $J_2$ (in K) of CuCrO$_4$ obtained from GGA+$U$ calculations with $U_{\rm eff}$ = 4 and 6 eV. The left column for each $U_{\rm eff}$ contains the theoretical results, while the values in the right column are the scaled theoretical results such that $J_{\rm nnn}$ equals the experimental finding, -27~K. The rightmost column summarizes the experimentally found spin exchange values. The final row show the Curie-Weiss temperatures of the scaled GGA+$U$ spin exchange parameters, calculated using the mean field expression;  $\theta_{\rm CW}$~=~$\frac{1}{3}\sum_{\substack{i}}z_iJ_{\rm{i}}S(S+1)$, where $z_i$ is the number of neighbor with which a single atom interacts with the spin exchange $J_i$, and the experimentally observed values (see below).}
\label{Table1}
\end{table}

The intra-chain spin exchanges $J_{\rm {nn}}$ and $J_{\rm {nnn}}$ are both afm and constitute the two dominant spin exchanges in CuCrO$_4$. The inter-chain parameter $J_2$, connecting Cu atoms related by a translation along $a$, is fm and, depending on the onsite repulsion parameter $U_{\rm eff}$, its magnitude amounts to 15 to 20\% of the intra-chain spin exchange $J_{\rm {nn}}$. $J_1$, which couples adjacent spin moments which are related by a translation along [110], is afm and comparatively small. Therefore, to a first approximation, CuCrO$_4$ can be described as a quasi 1D Heisenberg magnet with nn and nnn spin exchange interactions, both being afm. Since these 1D chains are connected by weak inter-chain exchanges ($J_{1}$ and $J_{2}$), long range ordering will eventually take place at low temperatures.

\subsection{Experimental}

A polycrystalline sample of CuCrO$_4$ was prepared by separately dissolving equimolar amounts of anhydrous Copper(II)acetate and Chromium(VI)oxide in distilled water, similar to the recipe given by  Arsene \emph{et al.}\cite{Arsene1978}. The two solutions were mixed and boiled to dryness. The resulting powder was heat treated in air at a temperature of 150$^{\rm {\circ}}$C for 2 days. The phase purity of the sample was checked by x-ray powder diffraction measurements using a STOE STADI-P diffractometer with monochromated Mo-$K$${\alpha_1}$ radiation. The powder pattern was analyzed using the Rietveld profile refinement method employed within the Fullprof Suite.\cite{Fullprof} No other reflections besides those of CuCrO$_4$ were observed.

Powder reflectance spectra of CuCrO$_4$ were collected at room temperature using a modified CARY 17 spectrophotometer,  equipped with an integrating sphere.  The spectrometer was operated in the single-beam mode using BaSO$_4$ as reflectance (white) standard.
CuCrO$_4$ powder was mixed with  BaSO$_4$ in a volumetric ratio CuCrO$_4$:BaSO$_4$ $\sim$1 : 5.

Temperature dependent electron paramagnetic resonance (EPR) spectra of a $\sim$~5~mg polycrystalline sample, contained within an EPR low-background suprasil$^\copyright$ quartz tube, were collected using $\sim$~9.5~GHz microwave radiation (Bruker ER040XK microwave X-band spectrometer,  Bruker BE25 magnet equipped with a BH15 field controller calibrated against Diphenylpicrylhydrazyl (DPPH)).

The molar magnetic susceptibilities, $\chi_{\rm mol}$, of a polycrystalline sample weighting $\sim$~84~mg  were measured with various fields between 2~K and 350~K using a SQUID magnetometer (MPMS, Quantum Design). The raw magnetization data were corrected for the magnetization of the sample container.

The specific heats, $C_p$,  of a powder sample weighting $\sim$~2.4~mg were determined as a function of the temperature and magnetic field with a relaxation-type calorimeter (PPMS, Quantum Design) for the temperature range 0.4~K to 50~K and magnetic fields up to 9~T.

The relative dielectric constant, $\epsilon_{\rm r}$, was measured at a constant frequency and excitation voltage, 1000~Hz and 15~V, respectively, with an Andeen-Hagerling 2700A capacitance bridge on a compacted powder (thickness: $\sim$~0.8~mm, $\varnothing$: 3.6~mm).

\subsection{Results and Discussion}

\begin{figure}[htp]
\includegraphics[width=8cm ]{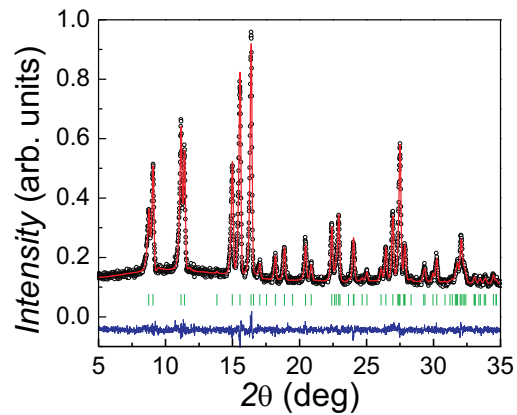}
\caption{(Color online) (o): Measured x-ray diffraction pattern of CuCrO$_4$ (wavelength 0.709 \AA\, Mo-$K$${\alpha_1}$ radiation). Solid (red) line: Fitted pattern ($R_p$~=~3.42~\%, reduced $\chi^2$~=~1.15) using the parameters given in Table \ref{Table2}. Solid (blue)  line (offset): Difference between measured and calculated patterns.
The positions of the Bragg reflections used to calculate the pattern are
marked by the (green) vertical bars in the lower part of the
figure.}
\label{Fig3}
\end{figure}

Figure \ref{Fig3} shows the measured and simulated x-ray powder diffraction patterns of the sample of CuCrO$_4$ used for all subsequent characterization. The refined atomic parameters and the lattice parameters are summarized in Table \ref{Table2} and were found to be in good agreement with the previously published single crystal results.\cite{Seferiadis1986}

\squeezetable
\begingroup
\begin{table}
\squeezetable
\centering
\begin{tabular}{ c  c c c c c }
\hline\hline
atom&Wyckoff site&x&y&z&$B_{\rm{iso}}$ (\AA $^2$)\\
\hline
Cu&4a&0&0&0&0.09( 8)\\ \\
Cr&4c&0&0.3700(3)&0.25&0.90( 8)\\ \\
O1&8f&0&0.2652(5)&0.0320(9)&0.80(12)\\ \\
O2&8g&0.2326(7)&-0.0198(6)&0.25&0.80(12)\\
\hline\hline
\end{tabular}
\caption{Atomic positional parameters of CuCrO$_4$ (SG: \textit{Cmcm}) as obtained from a profile refinement of the x-ray powder diffraction pattern, collected at room temperature. The lattice parameters amount to $a$~=~5.4388(5)~\AA, $b$~=~8.9723(8)~\AA\ and $c$~=~5.8904(6)~\AA.}\label{Table2}
\end{table}
\endgroup

\begin{figure}[!h]
\includegraphics[width=8cm ]{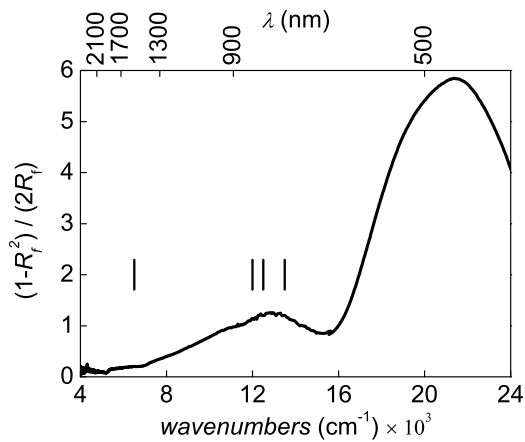}
\caption{(Color online) Powder reflectance spectrum of CuCrO$_4$. Black vertical bars mark the  ligand-field transition energies,  for the CuO$_6$ distorted octahedron, obtained from AOM calculations. We show the Kubelka-Munk relation, (1-$R_f$)$^2$/(2 $R_f$), where $R_f$ = $I$(CuCrO$_4$)/$I$(BaSO$_4$)
and $I$(CuCrO$_4$) and $I$(BaSO$_4$) are the reflected light intensities of the sample and the BaSO$_4$ standard, respectively.\cite{Kortum1969}}
\label{Opt}
\end{figure}

Figure \ref{Opt} displays the optical spectrum of CuCrO$_4$ which is consistent with the deep brownish-red color of the CuCrO$_4$ powder. The spectrum is dominated by a strong absorption band centered at 21500~cm$^{-1}$ (466 nm) which we attribute  to an O$^{2-}$~$\rightarrow$~Cr$^{6+}$ charge transfer transition, in agreement with observations for other hexavalent chromates.\cite{Johnson1970,Lever1984}
In the near infrared regime (NIR) the spectrum exhibits  a maximum at
$\tilde\nu_3$~=~13000 cm$^{-1}$ with a tail extending down to $\sim$~7000~cm$^{-1}$. Two subsequent faint shoulders are seen within the slope at
$\tilde\nu_2$~=~11000~cm$^{-1}$ and  $\tilde\nu_1 \sim$~8000~cm$^{-1}$.

Using ligand-field considerations (see below) the observed absorption bands ($\tilde\nu_1$, $\tilde\nu_2$ and $\tilde\nu_3$) can be assigned to   Cu$^{2+}$ $d$~-~$d$ transitions,
$^2B_{1g}$~$\rightarrow$~$^2A_{1g}$  ($z^2$~$\rightarrow$~$x^2$~-~$y^2$),
$^2B_{1g}$~$\rightarrow$~$^2B_{2g}$  ($xy$~$\rightarrow$~$x^2$~-~$y^2$), and
$^2B_{1g}$~$\rightarrow$~$^2E_{g}$  ($xz, yz$~$\rightarrow$~$x^2$~-~$y^2$), respectively.\cite{Jorgensen1963,Richardson1993,Larsen1974,Figgis2000}

From $\tilde\nu_1$, $\tilde\nu_2$, and $\tilde\nu_3$ the crystal field splitting, 10$Dq$ for CuO$_6$,  can be calculated using the relation,

\begin{equation*}
    10Dq=\tilde\nu_3-(\frac{\tilde\nu_3-\tilde\nu_2}{3})-(\frac{\tilde\nu_1}{2}),
\end{equation*}

which yields a value of
10$Dq$~$\sim$~8300~cm$^{-1}$. This value is similar to crystal field splitting values  previously reported e.g. for Cu$^{2+}$ aquo-complexes.\cite{Figgis2000}

UV/vis spectra for CuCrO$_4$ have been reported before by Baran and an assignment of the observed transitions has been been proposed.\cite{Baran1994}
Based on calculations within the framework of the angular overlap model (AOM)\cite{Jorgensen1963,Richardson1993,Larsen1974,Figgis2000} we argue that this assignment has to be revised.

Within the AOM model the pairwise interactions of the ligands with the $d$-orbitals are encoded into the parameters, $e_{\sigma}$, $e_{\pi,x}$ and $e_{\pi,y}$ which take care of interaction along and perpendicular to the Cu~-~O$_i$ ($i$~=~1, \ldots, 6) bond, respectively. The energies of the individual $d$-orbitals are obtained by summation over all  pairwise interactions.
The variation of the AOM parameters $e_{\sigma i}$ with the Cu~-~O$_i$ distance has been taken care of by,

\begin{equation*}
    e_{\sigma i}  \propto 1/r_i^n.
\end{equation*}

An exponent of $n\approx$~5 is derived from electrostatic and covalent theoretical bonding considerations.\cite{Kortum1969,Smith1969,Bermejo1983}
Measurements of the pressure dependence of $10Dq$ pointed to a similar exponent
5~$\leq$~$n$~$\leq$~6.\cite{Drickamer1973} For the sake of simplicity we have chosen $e_{\pi,x}$~=~$e_{\pi,y}$~=~1/4$e_{\sigma}$.  AOM calculations  have been performed using the  program CAMMAG.\cite{Gerloch1983,Cruse1980} Table \ref{AOMTable} summarizes the parameters which have been used for these calculations. The resulting  transition energies marked by vertical bars in Fig. \ref{Opt} are in good agreement with the centers of the experimentally observed absorption features.

\begin{table}[!h]
\begin{tabular}{ccc}\hline\hline
 & O$_{\rm eq}$  &O$_{\rm ax}$   \\
\hline
 $d$ (Cu-O) (\AA) & 1.965 (4$\times$) & 2.400 (2$\times$)\\ \\
 $e_{\sigma}$ (cm$^{-1}$)  & 5600 & 2061 \\ \\
  $e_{\pi,x}$ (cm$^{-1}$)  & 1400 & 515 \\ \\
    $e_{\pi,y}$ (cm$^{-1}$)  & 1400 & 515 \\
\hline\hline
\end{tabular}
\caption{Parameter used in the AOM calculations. The equatorial plane forms a rectangle with the equatorial O$_{\rm eq}$~-~Cu~-~O$_{\rm eq}$ bonds enclosing an $\angle$ of 81.92$^{\rm o}$ and 98.08$^{\rm o}$, respectively. The Racah parameters amounted to $B$~=~992~cm$^{-1}$, $C$~=~3770~cm$^{-1}$ yielding a ratio $C$/$B$~=~3.8, as given for the free Cu$^{2+}$ ion.\cite{Figgis2000} As for the aquo-complex the nephelauxetic ratio $\beta$ was chosen to be 0.80, and the spin-orbit coupling parameter $\zeta$~=~664~cm$^{-1}$ was reduced by 20\% as compared to the free ion value.\cite{Figgis2000,McClure1959}}\label{AOMTable}
\end{table}

In addition to the energy of the excited electronic states of the isolated CuO$_6$ unit, its magnetic properties are also obtained from the AOM calculations. The parametrization leads to an average $g_{\rm av}$ = 2.18 and a strongly anisotropic $g$-tensor with $g_x$~=~2.07, $g_y$~=~2.07, and $g_z$~=~2.39 along the principle axes. The $z$-direction of the $g$-tensor lies along   the Cu~-~O$_{\rm ax}$ bond direction.

\begin{figure}
\includegraphics[width=7.5cm]{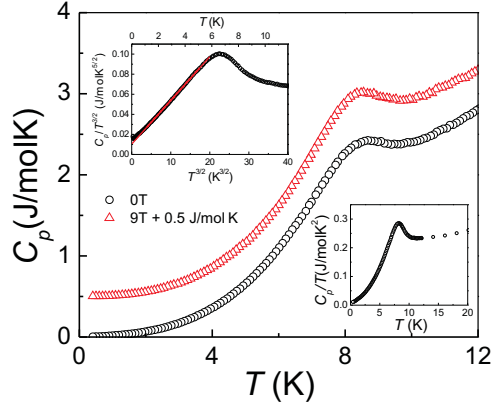}\\
\caption{(Color online) (Black) o and  (red) $\bigtriangleup$: Heat capacity of CuCrO$_4$ at 0~T and 9~T, respectively. The latter data have been shifted by +0.5 J/molK. Upper left inset: $C_p/T^{3/2}$ plotted versus $T^{3/2}$ to highlight the low-temperature $T^{3/2}$ power law. The (red) solid line is a fit of the data to eq. (\ref{eqcp}) with parameters given in the text.
Lower right inset: $C_p/T$ depicted against $T$ in the low-temperature regime.}\label{Fig4}
\end{figure}

The results of the specific heat measurements for magnetic fields of 0~T and 9~T are displayed in Fig \ref{Fig4}. The 0~T data reveal a rather broad, smeared, $\lambda$-type anomaly centered at 8.2(5)~K marking the onset of long-range magnetic ordering. Within experimental error the data measured in a magnetic field of 9~T are identical to those obtained at 0~T.
The plot of $C_p/T$ versus $T$ given in the low right inset of Fig. \ref{Fig4} enables the estimation  of the entropy contained within the anomaly, which equates to $\sim$~0.6~J/molK or $\sim$~10~$\%$ of the expected entropy of   a $S$ = 1/2 system, $R$\,ln(2), where $R$ is the molar gas constant. 90\% of the entropy has already been removed by short-range afm ordering above $T_{\rm N}$.

At low temperatures, the heat capacity comprises of a phonon and magnon contribution.
The temperature dependence of the phonon contributions to the heat capacity  can be described by a Debye-$T^3$ power law. The magnon heat capacity at low temperatures  varies with a power law depending on the spin wave dispersion relation and the dimensionality of the lattice.
For a three-dimensional (3D) magnetic lattice, one obtains a $T^3$ power law for afm magnons, and a $T^{3/2}$ power law for fm magnons.\cite{deJongh}
The $C_p/T^{3/2}$  versus $T^{3/2}$ plot shown in the upper left inset of Fig. \ref{Fig4} demonstrates that at low temperatures the heat capacity conforms well to a $T^{3/2}$ power law, with the coefficient of the fm magnon contribution given by the non-zero intercept with the ordinate, $\gamma$, according to,

\begin{equation}\label{eqcp}
C_p/T^{3/2} = \beta T^{3/2} + \gamma,
\end{equation}

where $\beta$ is related to the Debye temperature, $\theta_{\rm D}(0)$ at zero temperature  via,

\begin{equation}\label{eqDeb}
\beta = M R \frac{12\,\pi^4\,}{5}  (\frac{1}{\theta_{\rm D}})^3,
\end{equation}

with $M$ = 6  being the number of atoms per formula unit of CuCrO$_4$. While $\gamma$ can be expressed as,\cite{Martin1967}

\begin{equation}\label{eqgamma}
\gamma = A (\frac{k_{\rm B}}{J_{\perp}S})^{3/2}.
\end{equation}

Here we have assumed that the  Cu ribbon chains are coupled to neighboring chains by $J_{\perp}$, which we associate with the fm inter-chain spin exchange constant $J_2$ (see Table \ref{Table2}).

By using eq. (\ref{eqDeb}) we ascertain $\theta_D(T \rightarrow 0)$  to be,

\begin{equation*}\label{eqDeb1}
\theta(T \rightarrow 0) = 138(3)\,\,{\rm K},
\end{equation*}

and from the intercept and eq. (\ref{eqcp}) we obtain $\gamma$ as

\begin{equation*}\label{eqDeb2}
\gamma = 1.03(2) \times 10^{-2}\,\,\, {\rm J/molK^{5/2}}.
\end{equation*}

By using $J_{\perp} \sim J_2 \sim$ 12 K (see below) and $S$=1/2, the pre-factor $A$ in eq. (\ref{eqgamma}) amounts to

\begin{equation*}\label{eqDeb3}
A \approx 0.15\,\,\, {\rm J/molK}.
\end{equation*}

\begin{figure}
\includegraphics[width=7.5cm]{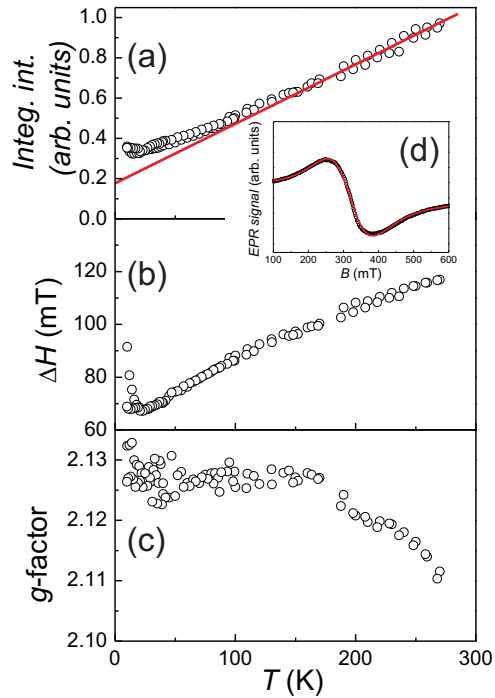}\\
\caption{(Color online) Results of the EPR measurements on a polycrystalline sample of CuCrO$_4$.
(a) (o) Inverse of the integrated intensity. The (red) solid line is a fit of eq. (\ref{eqIntInt}) to the high temperature data ($T \geq$~150~K).
(b) (o) The fitted half-width-at-half-maximum (HWHM) versus temperature.
(c) (o) $g$-factor versus temperature.
(d) (o)   EPR spectrum of CuCrO$_4$ measured at RT with $\sim$~9.45~GHz versus applied magnetic field.  The (red) solid line represents the fitted  derivative of a Lorentzian absorption line (eq. \ref{EPRpow}) to the measured spectrum.}\label{EPR}
\end{figure}

Figure \ref{EPR} summarizes the results of our  EPR measurements.
Near 3.4~kOe a single rather broad (peak-to-peak linewidth $\Delta H_{\rm pp} \approx$ 0.8 - 1 kOe) symmetric resonance line was observed. It can be well fitted to the derivative of a single Lorentzian absorption line  with a small contribution $|\alpha |\leq$ 0.04 of dispersion  according to

\begin{equation}\label{EPRpow}
\frac{dP_{\rm abs}}{dH} \propto \frac{d}{dH}\frac{\Delta H + \alpha(H - H_{\rm res})}{(H - H_{\rm res})^2 + \Delta H^2} + \frac{\Delta H + \alpha(H+H_{\rm res})}{(H + H_{\rm res})^2 + \Delta H^2}.
\end{equation}

As the linewidth (half-width at half-maximum (HWHM)), $\Delta H$, is of the same order of magnitude as the resonance field, $H_{\rm res}$ (see Fig. \ref{EPR}(a)), in eq. (\ref{EPRpow}) we took into account both circular components of the exciting
linearly polarized microwave field and therefore also included the
resonance at negative magnetic fields centered at $-H_{\rm res}$.

The resonance field of the room temperature  powder spectrum corresponds to a \emph{g}-factor of 2.117(2). Upon cooling a slight increase of the  \emph{g}-factor  with saturation to a value of $\sim$~2.125 below 150~K was observed (Fig. \ref{EPR}(c)).
Such a value  is somewhat lower than the expected average value $g_{\rm av}$ ascertained from the AOM calculations.
The resonance line is too broad to resolve the anisotropic \emph{g}-factors which range between $\sim$~2.39 and $\sim$~2.07 (see above).

The integrated intensity of the EPR resonance, $I(T)$  which is proportional to the  spin-susceptibility, increases with decreasing temperature down to $\sim$~15~K where a hump occurs. Above $\sim$~150~K,   $I(T)$ follows a Curie-Weiss type temperature-dependence,

\begin{equation}\label{eqIntInt}
I (T)\propto \frac{1}{T - \theta_{\rm EPR}},
\end{equation}

with
\begin{equation*}\label{eqIntTheta}
 \theta_{\rm EPR} \approx -60(5)\,\,{\rm K}.
\end{equation*}

The negative $T$-axis intercept indicates predominant afm spin exchange interactions. Deviations from the Curie-Weiss type temperature-dependence are ascribed to short-range afm correlations, which start to develop below $\sim$~150~K, similar to the behavior of the dc magnetic susceptibility (see below). The decrease of the integrated intensity below $\sim$~15~K signals the onset of long-range ordering.

The magnitude and temperature-dependence of the EPR linewidth, $\Delta H$, are similar to
those observed for the inorganic spin-Peierls system CuGeO$_3$ or the frustrated afm 1D system LiCuVO$_4$.\cite{Yamada1996,Krug2002}
The linewidth exhibits a concave temperature dependence with a linear increase at low temperatures and for $T \rightarrow \infty$  one  extrapolates a saturation value of $\sim$~1.4~kOe.

If we assume that the temperature-dependant  broadening
of the EPR resonance line is due to anisotropic or antisymmetric components in the exchange Hamiltonian,  the constant high-temperature value can be estimated from the Kubo-Tomita limit as,\cite{Kubo1954}
\begin{equation}\label{eqKT}
\Delta H (T \rightarrow \infty) \approx \frac{1}{g \mu_{\rm B}} \frac{\delta^2}{J},
\end{equation}

where $\delta$ indicates the deviations from the symmetric Heisenberg spin exchange and  \textit{J} is the afm    symmetric intrachain exchange.  If for CuCrO$_4$ we associate
$J$ with the nn spin exchange,  $\sim$~60~K (see below), we can estimate a  $\delta$ of $\sim$~3~K, i.e. 5\% of the symmetric exchange.

The linear slope of the linewidth at low temperatures can be explained using the formulism put forth by Oshikawa and Affleck\cite{Oshikawa1999,Oshikawa2002} predicting

\begin{equation}\label{eqKT2}
\Delta H (T) \propto \frac{\delta^2}{J^2} T.
\end{equation}

We find a linear slope, indicative of 1D afm system, of $\sim$~2.5~Oe/K, similar to that observed for CuGeO$_3$ ($\sim$~4.5~Oe/K).\cite{Yamada1996}

The magnetic susceptibility of a polycrystalline sample of CuCrO$_4$ was measured in magnetic fields of 1, 3, 5 and 7 Tesla. Above $\sim$~20~K the susceptibilities are independent of the magnetic field indicating negligible ferromagnetic impurities. The susceptibilities, $\chi_{\rm mol}(T)$, above $\sim$~150~K follow the modified Curie-Weiss law,

\begin{equation}\label{eq2}
\chi_{\rm mol}(T) = \frac{C}{T - \Theta} + \chi_{\rm dia} + \chi_{\rm VV}.
\end{equation}

$C$ is the Curie constant pertaining to the spin susceptibility of the Cu$^{2+}$ entities, \mbox{$C$ = $N_{\rm A}$$g^2$\,$\mu_{\rm B}^2$\,$S(S+1)$/3$k_{\rm B}$}. $\chi_{\rm dia}$ refers to the diamagnetic susceptibilities of the electrons in the closed shells, that can be estimated from the increments given by Selwood, which equates to -62$\times$10$^{-6}$ cm$^3$/mol.\cite{Selwood1956}

At high temperatures, $T \geq$~150~K, we fitted the molar susceptibility to the aforementioned modified Curie-Weiss law (eq. \ref{eq2}).
We found best agreement with the following parameters:

\begin{equation*}
g = 2.17(2)\,\,\,\,\,\, \rm{and} \,\,\,\,\,\, \theta = -56(1) K
\end{equation*}

and

\begin{equation*}
\chi_{\rm dia} + \chi_{\rm VV}\approx  +20\times 10^{-6}\,\,{\rm cm}^3/{\rm mol}.
\end{equation*}

This puts the Van Vleck contribution to $\approx$~+80$\times$ 10$^{-6}$cm$^3$/mol which is in reasonable agreement with what has been found for other Cu$^{2+}$ compounds (see Ref. \onlinecite{Banks2009} and refs. therein). The fitted $g$-factor is in good agreement with optical spectroscopy and the  Curie-Weiss temperature is negative and in accordance with $\theta_{\rm EPR}$.

Below 150~K there are deviations from the Curie-Weiss law attributed to increasing afm short range correlations. The susceptibility passes through a broad shoulder with a subsequent kink at $\sim$~8~K whereupon it becomes field dependent, with a tendency to diverge for small fields. With increasing  fields the divergence is suppressed and the kink becomes more apparent. By 7 T a pronounced rounded hump with a maximum at 14.2(2) K and a subsequent dip at 8.0(5) K become clearly visible.

In general, GGA+$U$ calculations overestimate the spin exchange constants typically by a factor up to 4, in our case 2. \cite{Koo2008a,Koo2008b,Xiang2007b}  By taking this into account and by using a mean field approach one calculates, from the spin exchange parameters summarized in Table \ref{Table2}, a (negative) Curie-Weiss temperatures ranging between -38~K to -45~K, consistent with the experimental observations.

Our GGA+$U$ calculations indicate that CuCrO$_4$ can be described by a Heisenberg 1D chain with afm nn and afm nnn spin exchanges, with significantly weak inter-chain interactions ($J_2$/$J_{\rm {nn}}$~$<$~0.19).
Therefore, we modeled the magnetic susceptibility of CuCrO$_4$ against exact diagonalization results for the susceptibility $\chi_{\rm chain}(g,\alpha,J_{\rm{nnn}})$ of a single chain provided by Heidrich-Meissner \textit{et al.},
\cite{Heidrich2006,Heidrichweb} with

\begin{equation}\label{eqalpha}
\alpha = J_{\rm {nn}}/J_{\rm {nnn}}.
\end{equation}

Interchain spin exchange is treated within a mean-field approach according to,
\cite{Carlin1986}

\begin{equation}\label{eqMF}
\chi_{\rm mol}(T) = \frac{\chi_{\rm chain}(T)}{1 - \lambda\,\chi_{\rm chain}(T)} + \chi_0.
\end{equation}

By using the already known values, $\chi_{\rm{0}}$ = $\chi_{\rm{dia}}$ + $\chi_{\rm{VV}}$ = +20$\times$10$^{-6}$\,\,cm$^3$/mol as found from the fit of the high temperature magnetic susceptibility and a $g$-factor of 2.13 obtained from the EPR measurements, the simulated results can be compared to experimental data. The mean-field parameter, $\lambda$, in eq. (\ref{eqMF}) can be ascribed to the inter-chain spin exchange interactions  according to\cite{Carlin1986}

\begin{equation}\label{eqlambda}
\lambda = (z_1\,J_1+z_2\,J_2)/N_{\rm A}g^2\mu_{\rm B}^2,
\end{equation}

wherein,  $z_1$~=~4 and $z_2$~=~2 count the number of spin moments with which  a chain spin interacts through the inter-chain spin exchange interactions, $J_1$ and $J_2$, respectively.
Guided by the GGA+$U$ results, the ratio $\alpha$ is positive and in the regime of 1.5 to 2.5. Within this range for $\alpha$ we find best agreement of our experimental data with the model calculations for,

\begin{equation*}\label{eqMFres2}
\alpha \approx 2, \,\,\,\,\,\, {\rm{implying}} \,\,\,\,\,\, J_{\rm {nnn}} = -27(2)~{\rm{K}},
\end{equation*}

and a positive $\lambda$, which amounts to

\begin{equation*}\label{eqMFres4}
\lambda = 7(1)\, {\rm mol}/{\rm cm^3}.
\end{equation*}

\begin{figure}
\includegraphics[width=7.5cm]{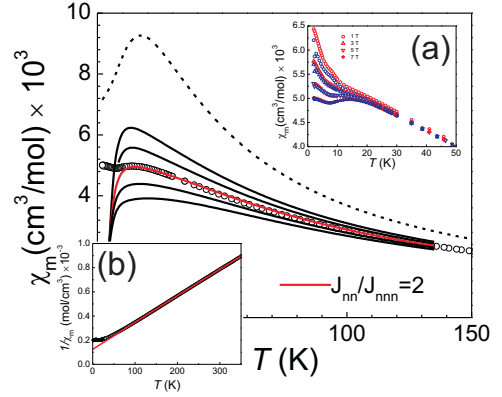}\\
\caption{(Color online) (main panel) (o) Temperature dependence of the molar magnetic susceptibility, $\chi$$_{\rm m}$, taken at 7~T. Colored solid lines  represent the exact diagonalization results by Heidrich-Meissner \textit{et al.} for various ratios  of  $J_{\rm nn}$/$J_{\rm nnn}$,  1.5, 1.75, 2 (red solid line), 2.25 and 2.5, from top to bottom, respectively. See text for more details.  The dashed line is the magnetic susceptibility of a $S$=1/2 Heisenberg chain with afm uniform nn spin exchange of -27~K.\cite{Johnston2000}
(a) red symbols: heating data, blue symbols: cooling data. $\chi_{\rm mol}$ versus temperature for various magnetic field. (b) (o) Reciprocal molar susceptibility versus temperature with a fit ((red) solid line) to a modified Curie-Weiss law (eq. (\ref{eq2})).}\label{Fig6}
\end{figure}

Figure \ref{Fig6} shows a comparison of the measured data and the mean-field corrected exact diagonalization results.

$\lambda >$ 0 indicates that the dominant inter-chain spin exchange is fm, consistent with our density functional calculations. The DFT calculations indicate $J_1$~$\approx$~-1/4$\times J_2$, irrespective of $U_{\rm eff}$. From eq. (\ref{eqlambda}) using $\lambda$~=~7(1)~mol/cm$^3$ we derive a value for $J_2$ which amounts to

 \begin{equation*}\label{eqMFres5}
J_2 = 12(2) {\rm K}.
\end{equation*}

This value is in good agreement with the scaled DFT result, see Table \ref{Table1}.

The inter-chain spin exchange can also be estimated from the N\'{e}el temperature, $T_{\rm N}$, which, according to the heat capacity data, amounts to (see above);

\begin{equation*}\label{eqTN}
T_{\rm N} \approx 8.2(5) {\rm K}.
\end{equation*}

Yasuda \textit{et al.} calculated the N\'{e}el temperature of a quasi 1D Heisenberg antiferromagnet on a cubic lattice with the isotropic inter-chain coupling $J_{\perp}$, inducing 3D long-range magnetic ordering at a N\'{e}el temperature, $T_{\rm N}$;\cite{Yasuda2005}

\begin{equation}\label{eq8Yasuda}
T_{\rm N}/|J_{\perp}| = 0.932\,\sqrt{ln(A) + \frac{1}{2}ln\,ln(A)},
\end{equation}

where $A$= 2.6\,$J_{\rm{\parallel}}$/$T_{\rm N}$ and $J_{\rm{\parallel}}$ is the intrachain spin exchange constant. If we assume $J_{\rm{\parallel}}$ to be our $J_{\rm {nn}}~\sim$~-60 K we find the inter-chain coupling to be

 \begin{equation*}\label{eqJperp}
|J_{\perp}| \approx 5 {\rm K},
\end{equation*}

consistent with the value obtained, from $\lambda$. The differences may arise, since our real system has two different inter-chain coupling constants, $J_1$ and $J_2$, as indicated by our density functional calculations.  Additionally, CuCrO$_4$ has a nnn spin exchange $J_{\rm {nnn}}$, which is not included in Yasuda's model.

\begin{figure}
\includegraphics[width=9cm]{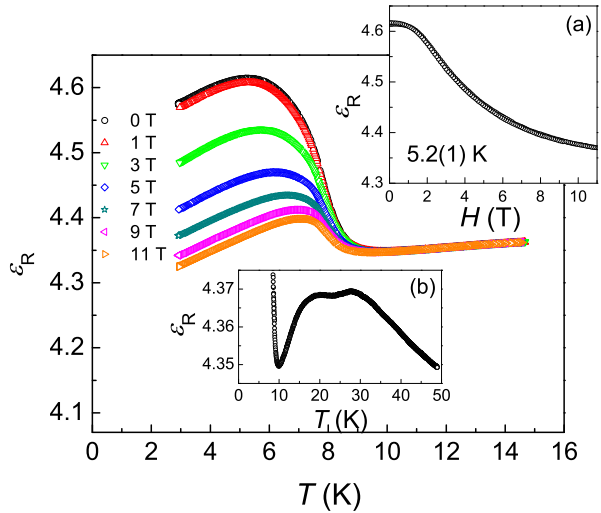}\\
\caption{
(Color online) (a) Colored symbols represent the relative dielectric constants, $\epsilon_{\rm r}$, versus temperature for different applied magnetic fields, as given in the legend. (b) The zero field relative dielectric constant is shown by the solid black line within a greater temperature range. (c) (o) The relative dielectric constant versus applied magnetic field  at a temperature of 5.2(1)~K.
}\label{epsilon}
\end{figure}

Figure \ref{epsilon} displays the temperature and magnetic field dependence of the relative dielectric constant, $\epsilon_{\rm r}$, of a compacted polycrystalline sample of CuCrO$_4$.

At room temperature, a value of $\sim$~48 was found for $\epsilon$$_{\rm r}$. With decreasing temperature, $\epsilon_{\rm r}$ is seen to decrease in a smooth fashion, until it passes through a shallow double maximum between 35 and 15 K, possibly indicating some magnetostriction induced by short range magnetic ordering processes above $T_{\rm N}$ (see Fig. \ref{epsilon} inset (b)). At 10~K a value of $\epsilon_{\rm}~\sim$~4.35 was measured.
Long-range magnetic ordering leads to a sizeable ME effect as evidenced in the $\epsilon_{\rm r}$, however, with a rather broad anomaly extending over the whole temperature range down to 3~K. Indication for a sharp spike near $T_{\rm N}$, as is frequently found in multiferroic systems, has not been seen. Similar broad anomalies, originating at $T_{\rm N}$, have been in seen in CuCl$_2$ and CuBr$_2$.\cite{Seki2010,Kremer2010}
In zero field a steep increase of $\epsilon_{\rm r}$ is seen to occur below $\sim$~8.5~K with a broad slightly asymmetric hump centered at $\sim$~5.35~K.
In zero field the increase of $\epsilon_{\rm r}$ from the paramagnetic phase to the maximum of the hump amounts to $\sim$~6\%. Applying a magnetic field decreases the ME anomaly and moves the maximum  to higher temperatures. The onset of the ME anomaly is not seen to move, in accordance with the aforementioned $C_p$ measurements (see Fig. \ref{epsilon}, inset (c)).
The decrease of $\epsilon_{\rm r}$ with a magnetic field starts above $ \sim$~1~T and tends to saturation at  sufficiently high fields.

In summary,
CuCrO$_4$ represents a new 1D quantum antiferromagnet with a remarkable pronounced ME anomaly below the N\'{e}el temperature of 8.2 K. Our density functional calculations indicate that, to a first approximation, the spin lattice of CuCrO$_4$ is a 1D Heisenberg chain with the unique situation that both, nn and nnn, spin exchanges are afm.
$J_{\rm {nn}}$/$J_{\rm {nnn}}$ is found to be close to 2, which places CuCrO$_4$  in the vicinity of the Majumdar-Ghosh point. The presence of sizeable ferromagnetic inter-chain spin exchange interaction leads to long-range magnetic ordering. The occurrence of the rather large ME anomaly below the N\'{e}el temperature is taken as evidence for non-collinear, possibly helicoidal, spin ordering in the 1D chains.  CuCrO$_4$ therefore represents a new interesting example for an unusual type-II multiferroicity system.
Neutron scattering investigations are scheduled to clarify the exact nature of the magnetic ground state of CuCrO$_4$.

\begin{acknowledgments}
The Authors would like to thank S. H\"ohn, E. Br$\ddot{\rm u}$cher and G. Siegle for experimental assistance and T. Dahmen for the sample preparation. Work at NCSU by the Office of Basic Energy Sciences, Division of Materials Sciences, U. S. Department of Energy, under Grant DE-FG02-86ER45259, and also by the computing resources of the NERSC center and the HPC center of NCSU.
\end{acknowledgments}

\bibliographystyle{apsrev}

\begin{thebibliography}{99}


%Introduction
\bibitem{Auciello1998}
O. Auciello, J. F. Scott, and R. Ramesh, Physics Today \textbf{51}, 22 (1998).

\bibitem{Spaldin2010}
N. A. Spaldin, S.-W. Cheong, and R. Ramesh, Physics Today \textbf{63}, 38 (2010).

\bibitem{Mostovoy2006}
M. Mostovoy, Phys. Rev. Lett. \textbf{96}, 067601 (2006).

\bibitem{Khomskii2009}
D. Khomskii, Physics \textbf{2}, 20 (2009).

\bibitem{Cheong2007}
S.-W. Cheong and M. Mostovoy, Nat. Mater. \textbf{6}, 13 (2007).

\bibitem{Tokura2010}
Y. Tokura and S. Seki,  Adv. Mater. \textbf{22}, 1554 (2010).

\bibitem{Brink2008}
J. van den Brink and  D. I. Khomskii, J. Phys.: Condens. Matter \textbf{20}, 434217 (2008).

\bibitem{Gibson2004}
B. J. Gibson, R. K. Kremer, A. V. Prokofiev, W. Assmus, and G.
J. McIntyre, Physica B \textbf{350}, e253 (2004).

\bibitem{Enderle2005}
M. Enderle, C. Mukherjee, B. F{\rm{\aa}}k, R. K. Kremer, J.-M.
Broto, H. Rosner, S.-L. Drechsler, J. Richter, J. Malek, A.
Prokofiev, W. Assmus, S. Pujol, J.-L. Raggazzoni, H.
Rakoto, M. Rheinst\"adter, and H. M. R${\rm \o}$nnow, Europhys.
Lett. \textbf{70}, 237 (2005).

\bibitem{Capogna2005}
L. Capogna, M. Mayr, P. Horsch, M. Raichle, R. K. Kremer, M. Sofin, A. Maljuk, M. Jansen, and B. Keimer, Phys. Rev. B \textbf{71}, 140402(R) (2005).


\bibitem{Kimura2008}
T. Kimura, Y. Sekio, H. Nakamura, T. Siegrist, and A. P. Ramirez, Nat. Mater.
\textbf{7}, 291 (2008).


\bibitem{Banks2009}
M. G. Banks, R. K. Kremer, C. Hoch, A. Simon, B. Ouladdiaf,
J.-M. Broto, H. Rakoto, C. Lee, and M.-H. Whangbo,
Phys. Rev. B \textbf{80}, 024404 (2009).


\bibitem{Capogna2010}
L. Capogna, M. Reehuis, A. Maljuk, R. K. Kremer, B. Ouladdiaf, M. Jansen, and B. Keimer,
Phys. Rev. B \textbf{82} 014407 (2010).

\bibitem{Mourigal2011}
M. Mourigal, M. Enderle, R. K. Kremer, J. M. Law, and B. F{\rm{\aa}}k,
Phys. Rev. B \textbf{83}, 100409(R) (2011).


\bibitem{Koo2011}
H.-J. Koo, C. Lee, M.-H. Whangbo, G. J. McIntyre, and R. K. Kremer, arXiv:1103.1616.


\bibitem{Katsura2005}
H. Katsura, N. Nagaosa, and A. V. Balatsky, Phys. Rev. Lett. \textbf{95}, 057205 (2005).

\bibitem{Sergienko2006}
I. A. Sergienko and E. Dagotto, Phys. Rev. B \textbf{73}, 094434 (2006).

\bibitem{Xiang2007}
H. J. Xiang and M.-H. Whangbo,
Phys. Rev. Lett. \textbf{99}, 257203 (2007).


\bibitem{Attfield1985}
J. P. Attfield, P. D. Battle, and A. K. Cheetham,
J. Solid State Chem.  \textbf{57}, 357 (1985).

\bibitem{Glaum1996}
R. Glaum, M Reehuis, N. St\"{u}{\ss}er, U. Kaiser, and F. Reinauer, J. Solid State Chem. \textbf{126}, 15 (1996).

\bibitem{Law2010}
J. M. Law, C. Hoch, M.-H. Whangbo, and R. K. Kremer,
Z. Anorg. Allg. Chem. \textbf{636}, 54 (2010).

\bibitem{Law2011}
J. M. Law, C. Hoch, R. Glaum, I. Heinmaa, R. Stern, M.-H. Whangbo, and R. K. Kremer,
submitted for publication.










\bibitem{Majumdar1969}
C. K. Majumdar and  D. K. Ghosh,  J. Math. Phys. \textbf{10}, 1388 (1969);
C. K. Majumdar and  D. K. Ghosh, J. Math. Phys. \textbf{10}, 1399 (1969).



\bibitem{Enderle2010}
M. Enderle,  B. F{\rm{\aa}}k, H.-J. Mikeska, R. K. Kremer,  A.
Prokofiev, and W. Assmus,  Phys. Rev. Lett. \textbf{105},  027003 (2010).



%Structure

\bibitem{Brandt1943}
K. Brandt, Arkiv f\"{o}r Kemi, Mineralogi Och Geologi. \textbf{17a}, 1 (1943).

\bibitem{Seferiadis1986}
N. Seferiadis and H. R. Oswald, Acta Cryst. \textbf{C43}, 10 (1986).




\bibitem{Whangbo2003}
M.-H. Whangbo, H.-J. Koo, and D. Dai, J. Solid State Chem. \textbf{176}, 417 (2003).

\bibitem{Dai}
 D. Dai and M.-H. Whangbo, J. Chem. Phys. \textbf{114}, 2887 (2001); D. Dai and M.-H. Whangbo, J. Chem. Phys. \textbf{118}, 29 (2003).


%DFT Spin Exchange Interactions



\bibitem{Kresse1993}
G. Kresse and J.  Hafner, Phys. Rev. B  \textbf{62}, 558 (1993).

\bibitem{Kresse1996a}
G. Kresse and J. Furthm\"uller, Comput. Mater. Sci. \textbf{6}, 15 (1996).

\bibitem{Kresse1996b}
G. Kresse and J. Furthm\"uller, Phys. Rev. B \textbf{54}, 11169 (1996).

%(2)
\bibitem{Perdew1996}
J. P. Perdew, S. Burke, and  M. Ernzerhof, Phys. Rev. Lett. \textbf{77}, 3865 (1996).

\bibitem{Dudarev1998}
S. L. Dudarev, G. A. Botton, S. Y. Savrasov, C. J. Humphreys, and A. P. Sutton, Phys. Rev. B \textbf{58}, 1505 (1998).



\bibitem{Koo2008a}
H.J. Koo and M.-H. Whangbo, Inorg. Chem. \textbf{47}, 128 (2008).

\bibitem{Koo2008b}
H.-J. Koo and M.-H. Whangbo,  Inorg. Chem. \textbf{47}, 4779 (2008).

\bibitem{Kang2009}
J. Kang, C. Lee, R. K.  Kremer, and M.-H. Whangbo, J. Phys.: Condens. Matter \textbf{21}, 392201 (2009).



\bibitem{Koo2010}
H.J. Koo and M.-H. Whangbo, Inorg. Chem. \textbf{49}, 9253 (2010).


\bibitem{Arsene1978}
J. Arsene, M. Lenglet, A. Erb, and P. Granger, Revue de Chimie Minerale \textbf{15}, 318 (1978).

\bibitem{Fullprof}
J. Rodr\'{\i}guez-Carvajal, Phys. B \textbf{55}, 192 (1993).

\bibitem{Johnson1970}
L. W. Johnson and S. P. McGlynn, Chem. Phys. Lett. \textbf{7}, 618 (1970).

\bibitem{Lever1984}
A. B. P. Lever, \textit{Inorganic Electronic Spectroscopy}, (Elsevier, Amsterdam 1984).


\bibitem{Kortum1969}
G. Kort\"um, \textit{Reflexionsspektroskopie}, (Springer-Verlag, Berlin 1969).

\bibitem{Figgis2000}
B. N. Figgis and M. A. Hitchman, \emph{Ligand field Theory and its Applications}, (Wiley-VCH, New York 2000).




\bibitem{Jorgensen1963}
C. K. J{\o}rgensen, R. Pappalardo, and H.-H. Schmidtke, J. Chem. Phys. \textbf{39}, 1422 (1963).

\bibitem{Richardson1993}
D. E. Richardson, J. Chem. Ed. \textbf{70}, 372 (1993).

\bibitem{Larsen1974}
E. Larsen and G. N. LaMar, J. Chem. Ed. \textbf{51}, 633 (1974).

\bibitem{Baran1994}
E. J. Baran, Spectrochim. Acta \textbf{50}A, 2385 (1994).




\bibitem{Smith1969}
D. W. Smith, J. Chem Phys. \textbf{50}, 2784 (1969).

\bibitem{Bermejo1983}
M. Bermejo and L. Pueyo, J. Chem. Phys. \textbf{78}, 854 {1983}.


\bibitem{Drickamer1973}
H. G. Drickamer, \emph{Electronic Transitions and the High Pressure Chemistry and Physics of Solids}, (Chapman and Hall, London 1973).



\bibitem{Gerloch1983}
M. Gerloch, \textit{Magnetism and Ligand Field Theory}, (Cambridge University Press, Cambridge 1983).

\bibitem{Cruse1980}
D. A. Cruse, J. E. Davies, J. H. Harding, M. Gerloch, D. J. Mackey, and R. F. McMeeking, CAMMAG: A Fortran Program, (Cambridge 1980).


\bibitem{McClure1959}
D. S. McClure, Solid State Physics  \textbf{9}, 399 (1959) (edt. by F. Seitz and D. Turnbull, Academic Press, New York, 1959).




%Heat Capacity


\bibitem{deJongh}
A. R. Miedema and L. J. deJongh, Adv. Phys. \textbf{23}, 1 (1974).

\bibitem{Martin1967}
D. H. Martin,  \emph{Magnetism in Solids}, (The Massachusetts Institute of Technology Press, Cambridge MA. 1967)


%Electron Spin Resonance


\bibitem{Yamada1996}
I. Yamada, M Nishi, and J Akimitsu,
J. Phys.: Cond. Matter \textbf{8}, 2625 (1996).

\bibitem{Krug2002}
H.-A. Krug von Nidda, L. E. Svistov, M. V. Eremin, R. M. Eremina, A. Loidl, V. Kataev, A. Validov,
A. Prokofiev, and W. Assmus, Phys. Rev. B \textbf{65}, 134445 (2002).


\bibitem{Kubo1954}
R. Kubo and K. Tomita,
J. Phys. Soc. Jpn. \textbf{9}, 888 (1954).

\bibitem{Oshikawa1999}
M. Oshikawa and I. Affleck,
Phys. Rev. Lett. \textbf{82}, 5136 (1999).

\bibitem{Oshikawa2002}
M. Oshikawa and I. Affleck,
Phys. Rev. \textbf{65}, 134410 (2002).


%Magnetic Susceptibility




\bibitem{Selwood1956}
P. W. Selwood, \textit{Magnetochemistry} 2nd ed. (Interscience, New York, 1956), p. 78.





\bibitem{Xiang2007b}
H. J. Xiang, C. Lee, and M.-H. Whangbo,  Phys. Rev. B  \textbf{76}, 220411(R) (2007).



\bibitem{Heidrich2006}
F. Heidrich-Meisner, A. Honecker, and T. Vekua,  Phys. Rev. B  \textbf{74}, 020403(R) (2006).


\bibitem{Heidrichweb}
For detailed numerical tables see
http://www.theorie.physik.uni-goettingen.de/~honecker/j1j2-td/.


\bibitem{Carlin1986}
R. L. Carlin, \textit{Magnetochemistry}, (Springer Berlin-Verlag 1986).

%\bibitem{Kluemper2000}
%A. Kl\"umper and D. C. Johnston,
%Phys. Rev. Lett \textbf{84}, 4701 (2000).

\bibitem{Johnston2000}
D. C. Johnston. R. K. Kremer, M. Troyer, X. Wang, A. Kl\"umper, S. L. Budko, A. F. Panchula, and P. C. Canfield,
Phys. Rev. B \textbf{61}, 9558 (2000).



\bibitem{Yasuda2005}
C. Yasuda, S. Todo, K. Hukushima, F. Alet, M. Keller, M. Troyer, and H. Takayama, Phys. Rev. Lett. \textbf{94}, 217201 (2005).


\bibitem{Seki2010}
S. Seki, T. Kurumaji, S. Ishiwata, H. Matsui, H. Murakawa, Y. Tokunaga, Y. Kaneko, T. Hasegawa, and
Y. Tokura, Phys. Rev B \textbf{82}, 064424 (2010).


\bibitem{Kremer2010}
R. K. Kremer, unpublished results.

\end{thebibliography}

\end{document}